\documentclass[twocolumn,showpacs,preprintnumbers,amsmath,amssymb,prl,floatfix]{revtex4}

\usepackage{graphicx}
\usepackage{dcolumn}
\usepackage{bm}

\begin{document}

\title{Band-Selective Filter in a Zigzag Graphene Nanoribbon}

\author{Jun Nakabayashi}
\email{nakabayashi@kh.phys.waseda.ac.jp}
\author{Daisuke Yamamoto}
\author{Susumu Kurihara}
\affiliation{
Department of Physics, Waseda University, Tokyo, Japan
}

\date{\today}

\begin{abstract}
Electric transport of a zigzag graphene nanoribbon through a step-like potential or a barrier potential is investigated by using the recursive Green's function method. In the case of the step-like potential, we demonstrate numerically that scattering processes obey a selection rule for the band indices when the number of zigzag chains is even; the electrons belonging to the ``even'' (``odd'') bands are scattered only into the even (odd) bands so that the parity of wavefunctions is preserved. The so-called valley-valve effect can be explained by this selection rule. In the case of the barrier potential, by tuning the barrier height to be an appropriate value, we show that it can work as the ``band-selective filter'', which transmits electrons selectively with respect to the indices of the bands to which the incident electrons belong. Finally, we suggest that this selection rule can be observed in the conductance by applying two barrier potentials.
\end{abstract}


\maketitle


Graphene, a single-layer graphite, is one of the most intriguing new materials, which has been studied vigorously since Novoselov {\it et al.} first succeeded in fabrication in 2004 \cite{graphene1}. Particularly, its excellent transport properties such as the high mobility at room temperature have attracted a lot of research interest to apply it in making future electronic devices \cite{graphene2,graphene3}. However, to realize graphene-based devices, there is a crucial problem: the characteristic gapless band structure makes the control of the electric current rather difficult. Many attempts to overcome the problem have been done so far, and gave rise to various benefical systems such as a graphene quantum dot \cite{graphene4}, epitaxial graphene on SiC substrates \cite{graphene5}, and a graphene nanoribbon \cite{band-structure}.

\begin{figure}
\includegraphics[scale=0.50]{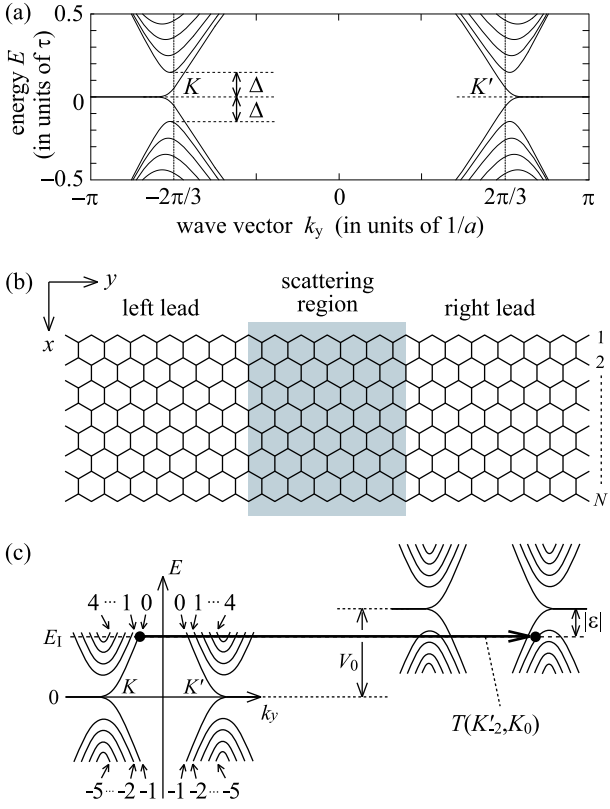}
\caption{\label{fig:1} (a) The band structure of a zigzag ribbon for the case of $N=30$. $2\Delta\simeq 3\tau\pi/N$ is the energy separation between the top of the next highest valence band and the bottom of the next lowest conduction band. (b) A schematic diagram of a zigzag ribbon. The electrostatic potential varies in the shaded region. The $x$ axis ($y$ axis) is taken to be perpendicular (parallel) to graphene lead lines. (c) A schematic diagram of the scattering process from $K_0$ into $K'_{-2}$ (see text). The bands are indexed as shown in the diagram.}
\end{figure}

We shall focus on a graphene nanoribbon with zigzag edges \cite{zigzag-ribbon} (hereafter we call it ``zigzag ribbon''). A band structure of a zigzag ribbon has two well-separated valleys (conically-shaped curves) $K$ and $K'$ around the vertices of the first Brillouin zone. In addition, the low-energy bands (the lowest conduction band and the highest valence band) are almost flat at the Fermi level due to the edge states \cite{band-structure, zigzag-ribbon}. Such a peculiar band structure has motivated much researchers to investigate the electric transport of a zigzag ribbon. Wakabayashi and Aoki \cite{current-blocking} found that the electric current is almost entirely blocked by a barrier potential when the incident energy $E_{\rm I}$ is in the range $[0,\Delta]$ while the barrier height $V_0$ is in the range $[\Delta,2\Delta]$ ($2\Delta$ is the energy separation between the top of the next highest valence band and the bottom of the next lowest conduction band; see Fig.~\ref{fig:1}(a)). This result is quite different from the case of bulk graphene, where incident electrons in the low-energy bands can pass a barrier potential of any height (known as the Klein Paradox \cite{klein-paradox}). At first, the polarization of two valleys was considered to be the origin of this current blocking effect, and then it was named ``valley-valve effect'' \cite{valley-valve} by using an analogy from the spin-valve effect \cite{spin-valve}. In the recent study \cite{theory-of-valley-valve}, however, Akhmerov ${\it et~al.}$ pointed out that the origin is not the valley polarization and showed that the behavior of the conductance is strongly connected with the parity of the number of zigzag chains $N$.

This striking current blocking effect has the possibility to control the electric current easily. In this letter, we investigate the role of a barrier potential applied to a zigzag ribbon in more detail, and suggest the way to control the current in a new light. The single-orbital tight-binding model is employed to describe the electronic states of a zigzag ribbon. The Hamiltonian is written as
\begin{eqnarray}
H=-\sum_{i,j}\tau_{ij}|i\rangle\langle j|+\sum_{i}V(y_i)|i\rangle\langle i|,
\end{eqnarray}
where the hopping integral $\tau_{ij}=\tau$ when $i$ and $j$ are nearest neighbor sites, and $\tau_{ij}=0$ otherwise. $V(y_i)$ is the electrostatic potential energy applied at the site $i$, which varies only along the direction of graphene lead lines (see Fig.~\ref{fig:1}(b)). The zero-bias conductance from the left lead to the right lead is given by the Landauer formula \cite{landauer}
\begin{eqnarray}
G=\frac{2e^2}{h}\sum_{\mu,\nu}T(\mu,\nu), \ \ \ T(\mu,\nu)=|t_{\mu\nu}|^2,\end{eqnarray}
where the summation runs over all incoming ($\nu$) and outgoing ($\mu$) channels, $t_{\mu\nu}$ is a transmission coefficient, and $T(\mu,\nu)$ represents a transmission probability. We specify channels $\nu$ and $\mu$ by the valley and band indices; the right-moving channel in the band $n$ belonging to the valley $K$ ($K'$) is denoted by $K_n$ ($K'_n$). For example, the transmission probability of the scattering process from the incoming channel in the band ``$0$'' belonging to the valley $K$ into the outgoing channel in the band ``$-2$'' belonging to the valley $K'$ is denoted by $T(K'_{-2},K_0)$ (see Fig.~\ref{fig:1}(c)). Hereafter we call bands with even (odd) indices ``even'' (``odd'') bands. For the calculation of the transmission coefficients, we adopt the recursive Green's function method developed by Ando \cite{conductance1,conductance2}.

\begin{figure}
\includegraphics[scale=0.70]{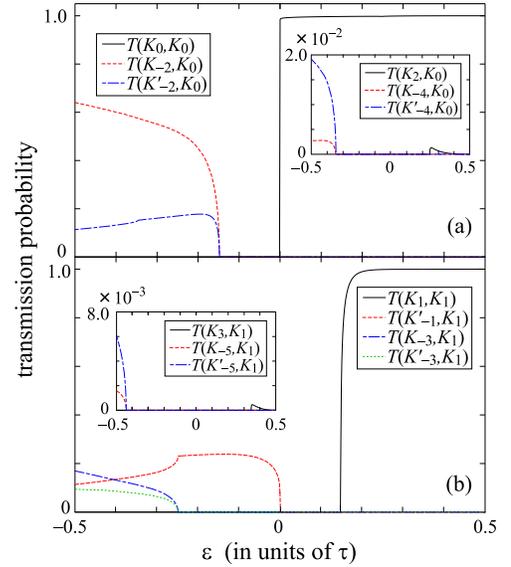}
\caption{\label{fig:2} (color online) The transmission probabilities for (a) $\nu=K_0$ and (b) $\nu=K_1$ with $N=30$. The insets are the enlargements of the low-probability regions. Although we do not show $T(K'_2,K_0)$, $T(K_4, K_0)$, $T(K'_4, K_0)$ and $T(K'_3,K_1)$ in the diagram, they have a very small yet non-zero value (less than $\sim 10^{-4}$), respectively. The others ($T(K_1,K_0)$, $T(K_2,K_1)$, etc.) are suppressed to 0.}
\end{figure}

{\it A step-like potential.}---Now, we consider the effect of a smooth step-like potential described as
\begin{eqnarray}
V(y)&=&V_0\Theta(y),\label{eq:step-like}\\
\Theta(y)&=&\left\{\begin{array}{ll}
0&(y<-d),\\
(1/2)\left[\sin(\pi y/2d)+1\right]&(|y|\leq d),\\
1&(y>d).\\
\end{array}\right.
\end{eqnarray}
In the present calculation, we set $d=10a$ ($a$ is the lattice constant) and the incident energy is fixed at $E_{\rm I}=0.5\tau$ while $V_0$ is varied from $0$ to $\tau$. The transmission probabilities are obtained as a function of $\varepsilon=E_{\rm I}-V_0$. From the results of the calculations, we find that scattering processes caused by the step-like potential obey the following selection rule: the electrons in the even (odd) bands can be scattered only into the even (odd) bands and other scattering processes are entirely restricted. For example, Figs.~\ref{fig:2}(a) and \ref{fig:2}(b) show the transmission probabilities for $\nu=K_0$ and $\nu=K_1$ with $N=30$. The electrons in the band $0$ are scattered only into the bands $0$, $\pm2$, and $\pm4$. Similarly, the electrons in the band $1$ are scattered only into the bands $\pm1$, $\pm3$, and $-5$.

\begin{figure}
\includegraphics[scale=0.65]{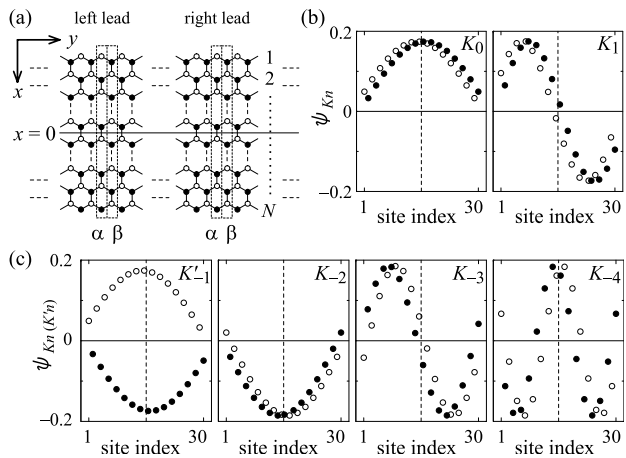}
\caption{\label{fig:3} (a) A schematic diagram of a zigzag ribbon for even $N$. Open and filled circles represent two inequivalent sublattices, respectively. One can regard a zigzag ribbon as a quasi-one-dimensional system. In this regime, the columns $\alpha$ and $\beta$ are treated as quasi sublattices. (b)(c) The wavefunctions can be written as $\Psi_{K_n(K'_n)}(x,y)=\psi_{K_n(K'_n)}(x){\rm e}^{{\rm i}k_y y}$. We show $\psi_{K_n(K'_n)}(x)$ of (b) the incoming channels $K_0$, $K_1$ on the quasi sublattice $\alpha$ in the left lead for $E_{\rm I}=0.5\tau$, (c) the outgoing channels $K'_{-1}$, $K_{-2}$, $K_{-3}$ and $K_{-4}$ on $\alpha$ in the right lead for $\varepsilon=-0.5\tau$. The wavefunctions on the quasi sublattice $\beta$ have the same parity as the one of the wavefunctions on $\alpha$. Here $N=30$.}
\end{figure}

The origin of this selection rule is the conservation of the parity of wavefunctions \cite{selection-rule, conservation-of-parity,andreev-reflection,footnote1,polyacene}. As is shown in Fig.~\ref{fig:3}(a), a zigzag ribbon has the reflection symmetry with respect to $x=0$ line when $N$ is even. Thus, wavefunctions must be either even or odd functions of $x$. For example, as is shown in Figs.~\ref{fig:3}(b) and \ref{fig:3}(c), the wavefunctions of the incoming channel $K_0$ and the outgoing channels $K_{-2}$ and $K_{-4}$ are even functions, while the incoming channel $K_1$ and the outgoing channels $K'_{-1}$ and $K_{-3}$ are odd functions. Similarly, other wavefunctions for even (odd) bands are even (odd) functions. Since the step-like potential given in Eq.~(\ref{eq:step-like}) cannot change the parity of wavefunctions, only the scattering processes which preserve the parity of wavefunctions can occur. Thus, the transmission probabilities such as $T(K_{-2},K_0)$ have non-zero values and the ones such as $T(K'_{-1},K_0)$ are restricted to $0$ (see Figs.~\ref{fig:2}(a) and \ref{fig:2}(b)). On the other hand, when $N$ is odd, a zigzag ribbon does not have the reflection symmetry and wavefunctions are neither even nor odd function of $x$. Consequently, the selection rule does not exist in this case.

The so-called valley-valve effect \cite{valley-valve} can be explained by this selection rule as follows. When $E_{\rm I}$ is in the range $[0,\Delta]$ while $V_0$ is in the range $[\Delta,2\Delta]$, the scattering processes from $K_0$ into $K'_{-1}$ are required for incident electrons to penetrate the step-like potential. However, when $N$ is even, the wavefunction of $K_0$ is an even function while the one of $K'_{-1}$ is an odd function. Thus, the scattering processes from $K_0$ into $K'_{-1}$ are entirely restricted because it does not preserve the parity of the wavefunction, and the current is blocked by the potential. When $N$ is odd, Since the selection rule does not exist, the current can flow through the potential \cite{theory-of-valley-valve}. The similar discussion on the valley-valve effect is shown in Ref.~\onlinecite{conservation-of-parity}.

{\it barrier potentials.}---By calculating the transmission probabilities individually, we have just shown that scattering processes obey the selection rule when $N$ is even. However, the existence of the selection rule shown in Figs.~\ref{fig:2}(a) and \ref{fig:2}(b) cannot be ascertained in experiments because the conductance is proportional to the sum of all transmission probabilities. Now, we show that the existence of the selection rule can be observed in the conductance by applying two barrier potentials.

\begin{figure}
\includegraphics[scale=0.55]{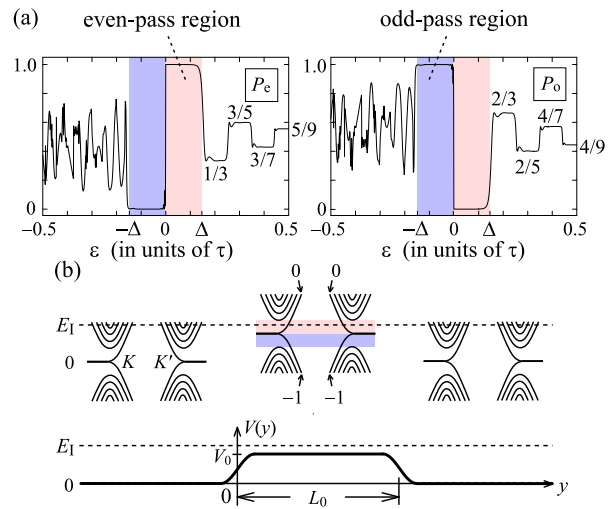}
\caption{\label{fig:4} (color online) (a) The results of the calculations of $P_{\rm e}$ and $P_{\rm o}$. The integral ratios shown in the top (bottom) panel represent the values of ${\cal N}_{\rm e}/{\cal N}$ (${\cal N}_{\rm o}/{\cal N}$) (see text). (b) Schematic diagrams of band structures and the barrier potential described as Eq.~(\ref{eq:barrier-like}). Here $L_0=50a$. The colored regions shown in the diagram correspond to the ones shown in (a).}
\end{figure}

First, we consider the case of one barrier potential
\begin{eqnarray}
V(y)=V_0[\Theta(y)-\Theta(y-L_0)].\label{eq:barrier-like}
\end{eqnarray}
To see the role of this potential, we introduce following quantities
\begin{eqnarray}
P_{{\rm e}}&=&\frac{\sum_{\mu_{\rm e},\nu}T(\mu_{\rm e},\nu)}{\sum_{\mu,\nu}T(\mu,\nu)},\\
P_{{\rm o}}&=&1-P_{{\rm e}}=\frac{\sum_{\mu_{\rm o},\nu}T(\mu_{\rm o},\nu)}{\sum_{\mu,\nu}T(\mu,\nu)},
\end{eqnarray}
where $\mu_{\rm e}$ ($\mu_{\rm o}$) represents the outgoing channels with even (odd) index. The results of the calculations are shown in Fig.~\ref{fig:4}(a). We can see at once that some plateaus appear when $\varepsilon>0$. In this region, the contributions of the transmission probabilities of the scattering processes between the same channels ($T(K_0,K_0)$, $T(K_1,K_1)$, etc.) are major ($\simeq 1$) and the others are minor (\hspace{0.3em}\raisebox{0.4ex}{$<$}\hspace{-0.75em}\raisebox{-.7ex}{$\sim$}\hspace{0.3em}$10^{-3}$) (see Fig.~\ref{fig:2}). Thus, $P_{\rm e}$ and $P_{\rm o}$ are given as $P_{\rm e}\simeq{\cal N}_{\rm e}/{\cal N}$ and $P_{\rm o}\simeq{\cal N}_{\rm o}/{\cal N}$, respectively. Here ${\cal N}_{\rm e}$ (${\cal N}_{\rm o}$) represents the number of the right-moving channels with even (odd) index in the potential region, and ${\cal N}={\cal N}_{\rm e}+{\cal N}_{\rm o}$.

Let us focus on the behavior in the colored regions shown in the diagram. In the red (or light gray) region (hereafter called the ``even-pass region''), $\varepsilon$ crosses only the band $0$ (see Fig.~\ref{fig:4}(b)), and thus $P_{\rm e}\simeq 1$ and $P_{\rm o}\simeq 0$ due to the selection rule \cite{footnote2}. In contrast, in the blue (or dark gray) region (hereafter called the ``odd-pass region''), $\varepsilon$ crosses only the band $-1$, and thus $P_{\rm o}\simeq 1$ and $P_{\rm e}\simeq 0$. These results mean that a barrier potential in a zigzag ribbon plays the role of a ``band-selective filter'': when $\varepsilon$ is tuned to be in the even- (odd-) pass region, the barrier potential transmits only the electrons in the even (odd) bands.

\begin{figure*}
\includegraphics[scale=0.70]{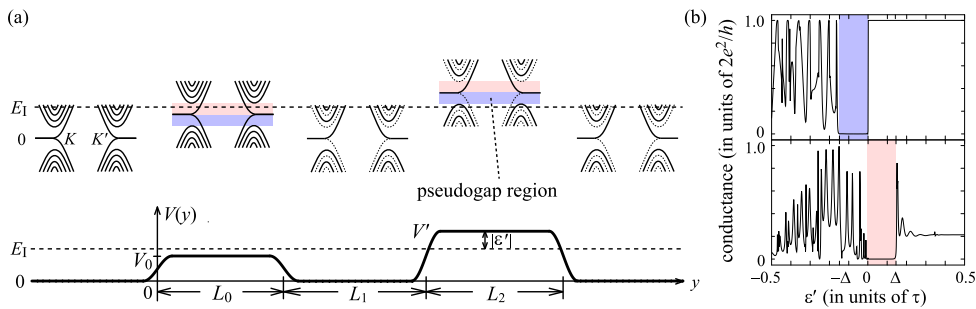}
\caption{\label{fig:5} (color online) (a) Schematic diagrams of band structures and the barrier potential described as Eq.~(\ref{eq:double-barrier}). (b) The results of the calculations of the conductances for $L_0=L_2=50a$ and $L_1=100a$. In the upper (lower) panel, $V_0$ is tuned to be $0.426\tau$ ($0.574\tau$) so that $\varepsilon$ is in the even- (odd-) pass region, and thus the odd- (even-) pass region in the second potential turns to the pseudogap region. In the region $\varepsilon'<0$, only interband scatterings occur, which makes the behavior of the conductances complicated.}
\end{figure*}

Next, we append one more barrier potential to the system as follows:
\begin{eqnarray}
V(y)\!\!&=&\!\!V_0[\Theta(y)-\Theta(y-L_0)]\nonumber\\
\!\!&+&\!\!V'[\Theta(y-L_0-L_1)\!-\!\Theta(y-L_0-L_1-L_2)],
\label{eq:double-barrier}
\end{eqnarray}
where $V_0$ is fixed so that the first barrier potential behaves as the band-selective filter (i.e. $|E_{\rm I}-V_0|\leq \Delta$), and $V'$ is varied from $0$ to $\tau$. Introducing two barrier potentials like that, one can observe the existence of the selection rule in the conductance. For example, when $\varepsilon$ is tuned to be in the even-pass region, the first potential works as the band-selective filter which transmits only the electrons in the even bands. In this case, since the incident electrons in the odd bands are reflected by this band-selective filter, all the electrons after passing the first potential belong to the even bands. Then, one can ignore the odd bands in the second potential because the scattering processes from even bands into odd bands are restricted by the selection rule. This indicates that the odd-pass region in the second potential turns to the pseudogap region (see Fig.~\ref{fig:5}(a); the negligible odd bands are shown as dashed lines). In fact, as is shown in the upper panel of Fig.~\ref{fig:5}(b), the conductance is entirely suppressed when $\varepsilon'=E_{\rm I}-V'$ is in the odd-pass (pseudogap) region. Similarly, when $\varepsilon$ is tuned to be in the odd-pass region, the even-pass region in the second potential turns to the pseudogap region (see the lower panel of Fig.~\ref{fig:5}(b)). Tuning the heights of two barrier potentials appropriately as above, one can handle the electric current easily, which will enables to apply a zigzag ribbon in making electronic devices.

In summary, we have studied electric transport of a zigzag ribbon through a step-like potential or a barrier potential by using the recursive Green's function method. It has been shown that scattering processes in a zigzag ribbon obey the selection rule for the band indices when the number of zigzag chains $N$ is even. The origin of the valley-valve effect is this selection rule. Moreover, we have also shown that a barrier potential can play the role of the band-selective filter, which transmits only the electrons in the bands with either even or odd index depending on $\varepsilon=E_{\rm I}-V_0$. Finally, we have suggested that the selection rule can be observed in the conductance by applying two barrier potentials to a graphene ribbon.

We thank Kenji Kamide for a useful discussion. One of the authors (D.Y.) is supported by a Grant-in-Aid from JSPS.


\begin{thebibliography}{9}
\bibitem{footnote2}
Actually the value of $P_{\rm e}$ ($P_{\rm o}$) is not strictly $1$ ($0$) because of the tunneling effect.
\end{thebibliography}
\end{document}